\documentclass[superscriptaddress,aps,prX,twocolumn,showpacs,nofootinbib,longbibliography,notitlepage,floatfix]{revtex4-2}
\usepackage[latin9]{inputenc}
\usepackage{mathtools}
\usepackage{amsbsy}
\usepackage{physics}
\usepackage{amstext}
\usepackage{xcolor}
\usepackage{listings}
\usepackage{array}

\usepackage{array}
\newcolumntype{C}[1]{>{\centering\arraybackslash}p{#1}}

\usepackage[unicode=true,pdfusetitle,
 bookmarks=false,
 breaklinks=false,pdfborder={0 0 1},backref=false,colorlinks=false]
 {hyperref}

\usepackage{etoolbox}
\usepackage[english]{babel}
\usepackage{chngcntr}
\hypersetup{
 bookmarksnumbered=false,bookmarksopen=false}
\makeatletter

\@ifundefined{textcolor}{}
{
 \definecolor{BLACK}{gray}{0}
 \definecolor{WHITE}{gray}{1}
 \definecolor{RED}{rgb}{1,0,0}
 \definecolor{BLUE}{rgb}{0,0,1}
 \definecolor{CYAN}{cmyk}{1,0,0,0}
 \definecolor{MAGENTA}{cmyk}{0,1,0,0}
 \definecolor{YELLOW}{cmyk}{0,0,1,0}
 \definecolor{ORANGE}{rgb}{1,0.5,0}
}

\usepackage{amsmath}
\usepackage{graphicx}
\usepackage{amssymb}
\usepackage{txfonts,color}
\makeatother

\definecolor{codegreen}{rgb}{0,0.6,0}
\definecolor{codegray}{rgb}{0.5,0.5,0.5}
\definecolor{codepurple}{rgb}{0.58,0,0.82}
\definecolor{backcolour}{rgb}{0.95,0.95,0.92}

\lstdefinestyle{mystyle}{
  backgroundcolor=\color{backcolour}, commentstyle=\color{codegreen},
  keywordstyle=\color{magenta},
  numberstyle=\tiny\color{codegray},
  stringstyle=\color{codepurple},
  basicstyle=\ttfamily\footnotesize,
  breakatwhitespace=false,         
  breaklines=true,                 
  captionpos=b,                    
  keepspaces=true,                 
  numbers=left,                    
  numbersep=5pt,                  
  showspaces=false,                
  showstringspaces=false,
  showtabs=false,                  
  tabsize=2
}

\begin{document}

\title{Automatic Detection of Nuclear Spins at Arbitrary Magnetic Fields via Signal-to-Image AI Model}

\author{B. Varona-Uriarte}
\affiliation{Department of Physical Chemistry, University of the Basque Country UPV/EHU, Apartado 644, 48080 Bilbao, Spain}
\affiliation{EHU Quantum Center, University of the Basque Country UPV/EHU, Leioa, Spain}

\author{C. Munuera-Javaloy}
\affiliation{Department of Physical Chemistry, University of the Basque Country UPV/EHU, Apartado 644, 48080 Bilbao, Spain}
\affiliation{EHU Quantum Center, University of the Basque Country UPV/EHU, Leioa, Spain}

\author{E. Terradillos}
\affiliation{TECNALIA, Basque Research and Technology Alliance (BRTA), Bizkaia Science and Technology Park, Astondo Bidea, Edificio 700, 48160 Derio, Spain}

\author{Y. Ban}
\affiliation{Departamento de F\'isica, Universidad Carlos III de Madrid, Avda. de la Universidad 30, 28911 Legan\'es, Spain}
\affiliation{TECNALIA, Basque Research and Technology Alliance (BRTA), Bizkaia Science and Technology Park, Astondo Bidea, Edificio 700, 48160 Derio, Spain}

\author{A. Alvarez-Gila}
\affiliation{TECNALIA, Basque Research and Technology Alliance (BRTA), Bizkaia Science and Technology Park, Astondo Bidea, Edificio 700, 48160 Derio, Spain}

\author{E. Garrote}
\affiliation{TECNALIA, Basque Research and Technology Alliance (BRTA), Bizkaia Science and Technology Park, Astondo Bidea, Edificio 700, 48160 Derio, Spain}
\affiliation{Department of Automatic Control and Systems Engineering, University of the Basque Country UPV/EHU, 48013 Bilbao, Spain}

\author{J. Casanova}
\affiliation{Department of Physical Chemistry, University of the Basque Country UPV/EHU, Apartado 644, 48080 Bilbao, Spain}
\affiliation{EHU Quantum Center, University of the Basque Country UPV/EHU, Leioa, Spain}

\begin{abstract}
Quantum sensors leverage matter's quantum properties to enable measurements with unprecedented spatial and spectral resolution. Among these sensors, those utilizing nitrogen-vacancy (NV) centers in diamond offer the distinct advantage of operating at room temperature. Nevertheless, signals received from NV centers are often complex, making interpretation challenging. This is especially relevant in low magnetic field scenarios, where standard approximations for modeling the system fail. Additionally, NV signals feature a prominent noise component. In this Letter, we present a signal-to-image deep learning model capable of automatically inferring the number of nuclear spins surrounding a NV sensor and the hyperfine couplings between the sensor and the nuclear spins. Our model is trained to operate effectively across various magnetic field scenarios, requires no prior knowledge of the involved nuclei, and is designed to handle noisy signals, leading to fast characterization of nuclear environments in real experimental conditions. With detailed numerical simulations, we test the performance of our model in scenarios involving varying numbers of nuclei, achieving an average error of less than $2\ \rm{kHz}$ in the estimated hyperfine constants. 
\end{abstract}
\maketitle

\emph{Introduction.--}  Fast and precise characterization of quantum registers is a central issue in the realm of quantum technologies, with vast applications in communication, computing, sensing, and simulation~\cite{gebhart23}. In particular, the efficient analysis of systems that comprise electronlike defects and nuclear spins in solid-state materials is pivotal for advancing quantum networks and quantum information processing~\cite{awschalom18, zwanenburg13}. In this context, deep learning-based approaches offer valuable solutions for addressing the challenges associated with quantum characterization~\cite{jung21, chen22, ban21}.

From a more technological perspective, nuclear-spin detection has experienced significant advancements with the use of NV centers in diamond~\cite{bradley19, vorobyov22}. NVs exhibit long coherence times~\cite{abobeih18} at room temperature \cite{nomura23, ohashi13}, which make them well-suited for a wide range of applications in the field of biological analysis~\cite{wu16, allert22, zhang21}. Additionally, NV-based quantum sensors are easy to initialize and readout by optical means~\cite{jelezko04, aharonovich09, dobrovitski13}, while their hyperfine levels can be coherently manipulated using microwave (MW) radiation~\cite{tamarat08, neumann08}. Regarding the NV environment, naturally $1.1\ \%$ of carbon nuclei are $^{13}\rm{C}$ featuring a spin-$\frac{1}{2}$~\cite{nier50}. Consequently, NVs couple to each $^{13}\rm{C}$ in the diamond lattice through a hyperfine vector $\vec{A}$ typically underlying dipolar interactions. Hence, rapid and precise characterization of each hyperfine vector would significantly contribute to assessing the potential of specific \textit{quantum nodes}, consisting of NVs and nuclei, for distinct quantum information processing tasks.

In this Letter, we present the SALI (Signal-to-image ArtificiaL Intelligence) model, able to characterize quantum nodes by identifying the number of involved nuclear spins and accurately estimating the hyperfine parameters of each NV-nucleus interaction. SALI comprises a 1D~$\rightarrow$~2D Convolutional Neural Network (CNN) module for processing 1D string data of NV measurements into a 2D image output, along with an image post-processing module. It is a compact model that offers automatic characterization (meaning it operates as a black box, without requiring human intervention) of nuclei in proximity to a probe NV, achieving the processing task within milliseconds, proving an advantage over classical algorithms \cite{taminiau12, oh20}. Notably, our model (i) exhibits high accuracy in predicting the value of hyperfine vectors over a wide range of values, (ii) effectively handles noisy signals commonly encountered in experimental scenarios, (iii) does not require prior knowledge of the number of nuclei in each node, and more importantly, (iv) performs well in low-field conditions, where conventional approximations break down, resulting in intricate and challenging signals. Through detailed numerical simulations, we evaluate the performance of our model in nodes containing up to 20 nuclei, considering both high and low magnetic field scenarios.

\emph{The system.--} We consider a quantum node consisting of a NV and $n$ $^{13}\rm{C}$ nearby nuclear spins with Larmor frequency $\omega_{\rm{L}}=\gamma_{\rm{n}}B_z$, such that $\gamma_{\rm{n}} = (2\pi) \times 10.705$ MHz/T, while the magnetic field $B_z$ is aligned with the NV axis ($\hat{z}$). The Hamiltonian that describes this system is 

\begin{equation}
H=\sum_{j=1}^n \omega_j \ \hat{\omega}_j \cdot\vec{I}_j + \frac{f(t)}{2}\sigma_z \sum_{j=1}^n \vec{A}_j \cdot \vec{I}_j,
\label{Hi}
\end{equation}

\begin{figure*}[t]
\includegraphics[width=1 \linewidth]{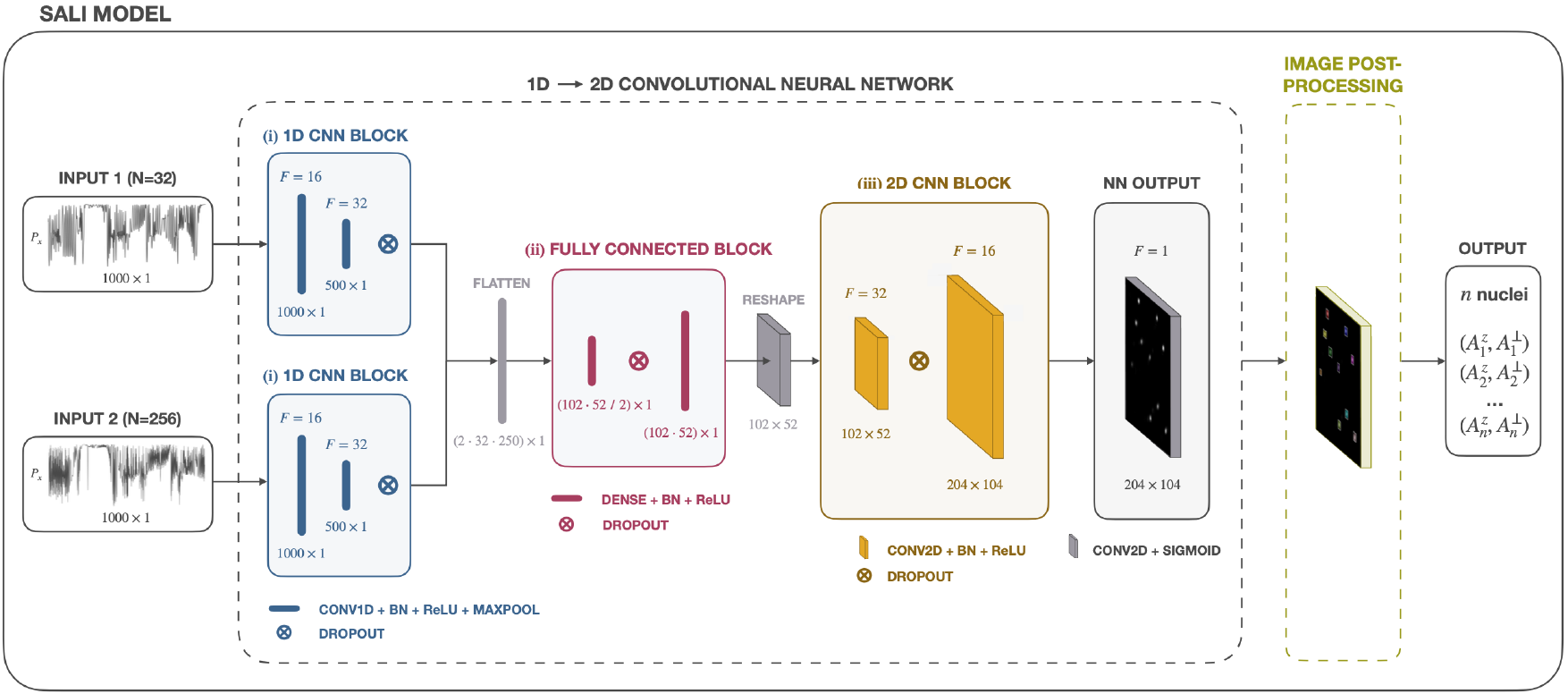}
\caption{Schematic representation of the SALI model. The neural network takes two input signals ($P_x$) obtained from CPMG sequences of different duration. In block (i) of the neural network, each input signal is independently processed through two 1-dimensional convolutional blocks.  Each block comprises two convolutional layers with $F$ filters of kernel size 3, followed by a batch normalization (BN) layer, a ReLU activation function, and a max-pooling layer with a window size of 2. The flattened outputs from these convolutional blocks are then concatenated and connected to block (ii), a fully connected block. The last dense layer in this block is reshaped into a rectangular image. Within block (iii), the second convolutional layer is a transposed convolutional layer with a stride of 2, effectively doubling the image's width and height. The 2-dimensional convolutional layers utilize $F$ filters with a kernel size of (3, 3). Block (iii) is subsequently connected to the output convolutional layer of the neural network, which is followed by an image post-processing module to extract the number of nuclei ($n$) within the sample and the corresponding coupling constant pairs $(A_j^z, A_j^{\bot})$. See in-depth explanation in SM \cite{supplemental}.}
\label{nn}
\end{figure*}

\noindent where $\omega_j \ \hat{\omega}_j = \gamma_{\rm{n}} B_z \  \hat{z} + \frac{1}{2} \vec{A}_j$, $\vec{A}_j=(A_j^{z}, A_j^{\bot})$ is the hyperfine vector joining the NV with the $j$th nucleus with spin operator $\vec{I}_j$, and $f(t)=\pm 1$ is the modulation function that appears as a consequence of the introduced MW driving (for more details see Supplemental Material (SM)~\cite{supplemental}). In particular, we consider trains of $\pi$-pulses over the NV (which is initialized to the $|+\rangle$ state, such that  $\sigma_x|+\rangle = |+\rangle $) according to the CPMG sequence \cite{carr54, meiboom58}, after which the NV is measured.  Repeating this process $\rm{N}_{\rm{m}}$ times, one estimates the survival probability ($P_x$) of the initial state $|+\rangle$. In the ideal scenario of infinite number of measurements (note our numerical simulations consider a finite number of measurements and decoherence effects) $P_x$ reads

\begin{equation}
P_x = \frac{1}{2} \left(1+\displaystyle\prod_{j=1}^n M_j\right),
\label{probability}
\end{equation}
where 
\begin{equation}
M_j = 1-m_{j,x}^2 \frac{(1-\cos{\alpha_j})(1-\cos{\beta})}{1+\cos{\alpha_j}\cos{\beta}-m_{j,z}\sin{\alpha_j}\sin{\beta}}\sin{\frac{N\phi_j}{2}}^2 ,
\label{mj}
\end{equation}

\begin{equation}
\cos{\phi_j} = \cos{\alpha_j}\cos{\beta}-m_{j,z}\sin{\alpha_j}\sin{\beta},
\label{cos}
\end{equation} 

\noindent with $m_{j,z} = \frac{(A_j^{z}+\omega_L)}{\tilde{\omega}_j}$, $m_{j,x} = \frac{A_j^{\bot}}{\tilde{\omega}_j}$, $\tilde{\omega}_j = \sqrt{(A_j^{z}+\omega_L)^2+{A_j^{\bot}}^2}$, $\alpha_j = \tilde{\omega}_j\tau$, $\beta=\omega_L \tau$, and $\tau$ is half the interpulse spacing of a CPMG sequence (see full derivation in Supplemental Material from Ref.~\cite{taminiau12}). In a scenario such that  $\omega_{\rm{L}}\gg A_j^z, A_j^{\bot}$ (namely, at high magnetic field, $B_z = 0.056 \rm{T}$ in our case) $P_x$ exhibits clear resonance peaks at  $\tau = \frac{k\pi}{2\omega_j}$. In this context, techniques based on classical algorithms \cite{taminiau12, oh20}, as well as deep learning models~\cite{jung21}, are used to find $A_j^{z}$ and $A_j^{\bot}$. However, in the low-field regime, where the condition $\omega_{\rm{L}}\gg A_j^z, A_j^{\bot}$ does not hold ($B_z = 0.0056 \rm{T}$ in our particular case), resonance peaks  cannot be observed~\cite{supplemental}. More specifically, in this regime $P_x$ shows an intricate behavior that makes previously mentioned techniques for system characterization challenging.

Now we introduce our SALI model designed to effectively process complex signals across diverse magnetic field scenarios, leading to $(A_j^{z}, A_j^{\bot})$ as output. Our model showcases robust performance in handling noisy signals commonly encountered in experimental scenarios and operates seamlessly without requiring any prior information about the number of nuclei involved in the node.

\begin{figure*}
\includegraphics[width=1 \linewidth]{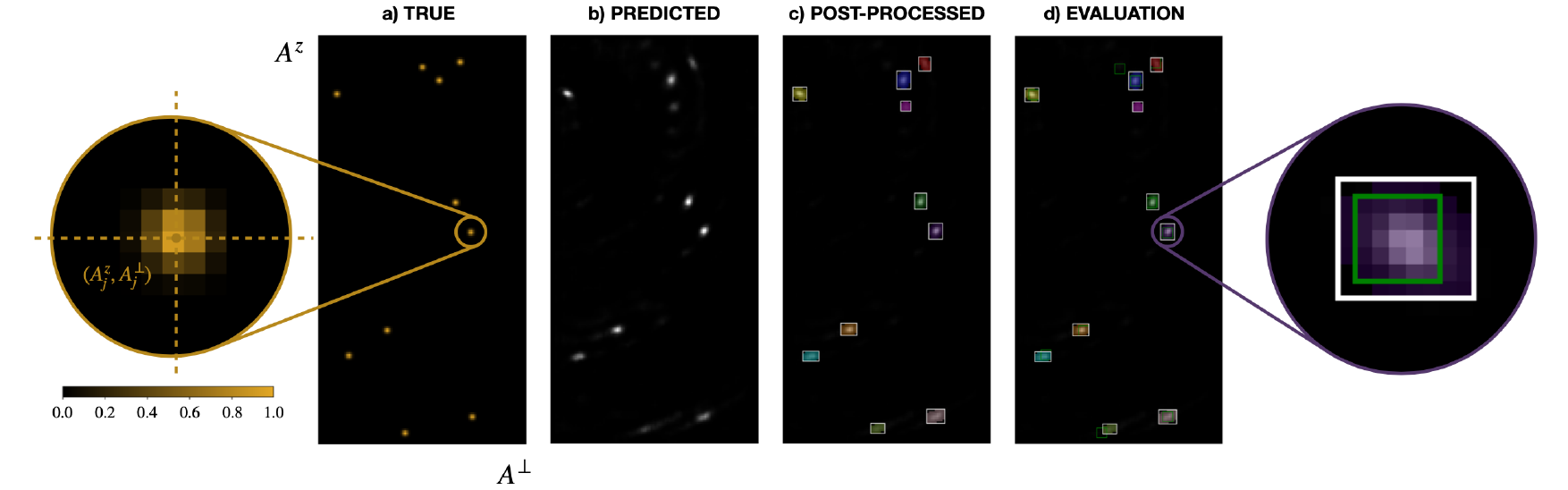}
\caption{(a) True output image of the low-field neural network. 10 nuclei with random coupling constants in the ranges $A^{z}\in [-100, \ 100] \ \rm{kHz}$ and $A^{\bot}\in [2, \ 102] \ \rm{kHz}$ are represented. Each nucleus appears as a $5\times5$ pixel region in the image, as depicted in the highlighted circular area. The pixel values in this region are chosen to match a Gaussian distribution centered in $(A_j^{z}, A_j^{\bot})$. (b) Output image predicted by the neural network. (c) Output image after the image post-processing. The white boxes represent predicted nuclei. (d) Evaluation of the model performance. The predicted nuclei (white boxes) are compared to the true nuclei (green boxes). In this example, we observe 9 TPs (true positives, i.e. correctly detected nuclei), 1 FP (false positive, i.e. predicted non-existent nucleus), and 1 FN (false negative, i.e. non-predicted nucleus). In the highlighted circular area we examine a specific TP nucleus. The $5\times 5$ green box bounds the nucleus in the true output, while the white box bounds the nucleus in the predicted output.}
\label{pixel}
\end{figure*}

\emph{The SALI model.--} A scheme of SALI is given  in Fig.~\ref{nn} (see caption for details on the architecture of the model). The 1D~$\rightarrow$~2D CNN module takes two input signals ($P_x$) coming from CPMG sequences with different number of pulses ($\rm{N}=32$ and $\rm{N}=256$ in our specific example). This approach ensures that each sequence exhibits different evolution times, enabling the network to infer both weak and strongly coupled nuclei. In this instance, these two specific sequences yielded highly favorable results. The architecture of the neural network is as follows: (i) Two separate 1D CNN blocks analyze the inputs. After processing the signals, the outputs of these blocks are flattened and concatenated into a single array. (ii) A fully connected block is introduced between the 1D CNN block and the next 2D CNN block, serving as an intermediary between these two blocks, and additionally, allowing for the adaptation of the final output image size. The outcome of the fully connected block is reshaped into a 2-dimensional array, treated as an image to exploit the spatial relations among adjacent pixels. (iii) In the 2D CNN block, the reshaped image is processed. Finally, this block connects to the output layer of the neural network (NN OUTPUT in Fig.~\ref{nn}), which is a convolutional layer with sigmoid activation function that encodes the target parameters $(A_j^{z}, A_j^{\bot})$ in a 2-dimensional image. 

For the training, validation, and testing of the neural network, we generated two distinct datasets, one for the high magnetic field scenario and the other for the low magnetic field scenario,  each comprising 3,600,000 samples. Each sample within these datasets contains a random number of nuclei ranging from 1 to 20. Each nucleus is characterized by random values of $A^{z}$ and $A^{\bot}$, falling within the ranges $A^{z}\in [-100, \ 100] \ \rm{kHz}$ and $A^{\bot}\in [2, \ 102] \ \rm{kHz}$, resulting in a set of coupling constants $(A_j^z, A_j^{\bot})$. Note that  Eqs.~(\ref{probability}-\ref{cos}) dictate that $P_x$ is symmetric with respect to the change $A^{\bot} \rightarrow -A^{\bot}$. Consequently, we only consider positive values for $A^{\bot}$. The input data strings ($P_x$) are generated with $\rm{N}=32$ $\pi$-pulses, with $\tau$ varying in the range $\tau_{32}\in[6,\ 50] \ \rm{\mu s}$, and $\rm{N}=256$ $\pi$-pulses, with $\tau$ in the range $\tau_{256}\in[10,\ 40] \ \rm{\mu s}$. Each $P_x$ contains $\rm{N}_{\rm{p}}=1000$ points, resulting in resolutions of $\Delta t_{32} = 44\ \rm{ns}$ and $\Delta t_{256} = 30\ \rm{ns}$. Thus, the employed CPMG sequences would last some milliseconds. We consider experimental conditions akin to those in~\cite{vorobyov22}. This is, we model potential decoherence effects over the NV sensor by adding an exponential factor $e^{-\tau/T_2}$ to $P_x$ with $T_2=200\ \rm{\mu s}$. Furthermore, we consider shot-noise by computing each average value in $P_x$ after simulating $\rm{N}_{\rm{m}}=1000$~measurements (see Sec. \ref{robustnesssection} in~\cite{supplemental} for evaluation of the model's robustness against shot-noise). The model parameters should be tailored to match the specifics of the experimental setup to ensure its optimal performance. In addition, the model could be fine-tuned with real experimental data (see \cite{mahmood18, too19, aggarwal22} for fine-tuning examples).

To supervise the training and validation stages, as well as for evaluation in the testing, nuclei are depicted in the true output image of the neural network as exemplified in Fig.~\ref{pixel}(a). This portrayal involves a fuzzy logic approximation through the use of Gaussians, reminiscent of representing a point in an image. As illustrated in Fig.~\ref{pixel}(a), each nucleus is depicted within a $5 \times 5$ pixel region centered around the nearest pixel to the true values $(A_j^{z}, A_j^{\bot})$. The loss function of the neural network is the Mean Squared Error (MSE) calculated across all pixel values within the output image.

Following the training process, the predicted output of the neural network exhibits distinct clusters of pixels, ideally resembling Gaussian functions see [Fig. \ref{pixel}(b)]. The next step is to analyze these clusters through the image post-processing module. This process begins with the application of erosion and dilation techniques, which are employed to smooth the image. Next, we perform a thresholding of the smoothed image and group the clusters of adjacent pixels with a connectivity routine. Finally, an area filter is applied to determine which predicted clusters qualify as nuclei [see white boxes in Fig.~\ref{pixel}(c)]. The result is the prediction of the number of nuclei $n$ and the corresponding coupling constant pairs $(A_j^{z}, A_j^{\bot})$, which are determined by the centroids of these clusters.

\begin{figure*}[t]
\includegraphics[width= 1 \linewidth]{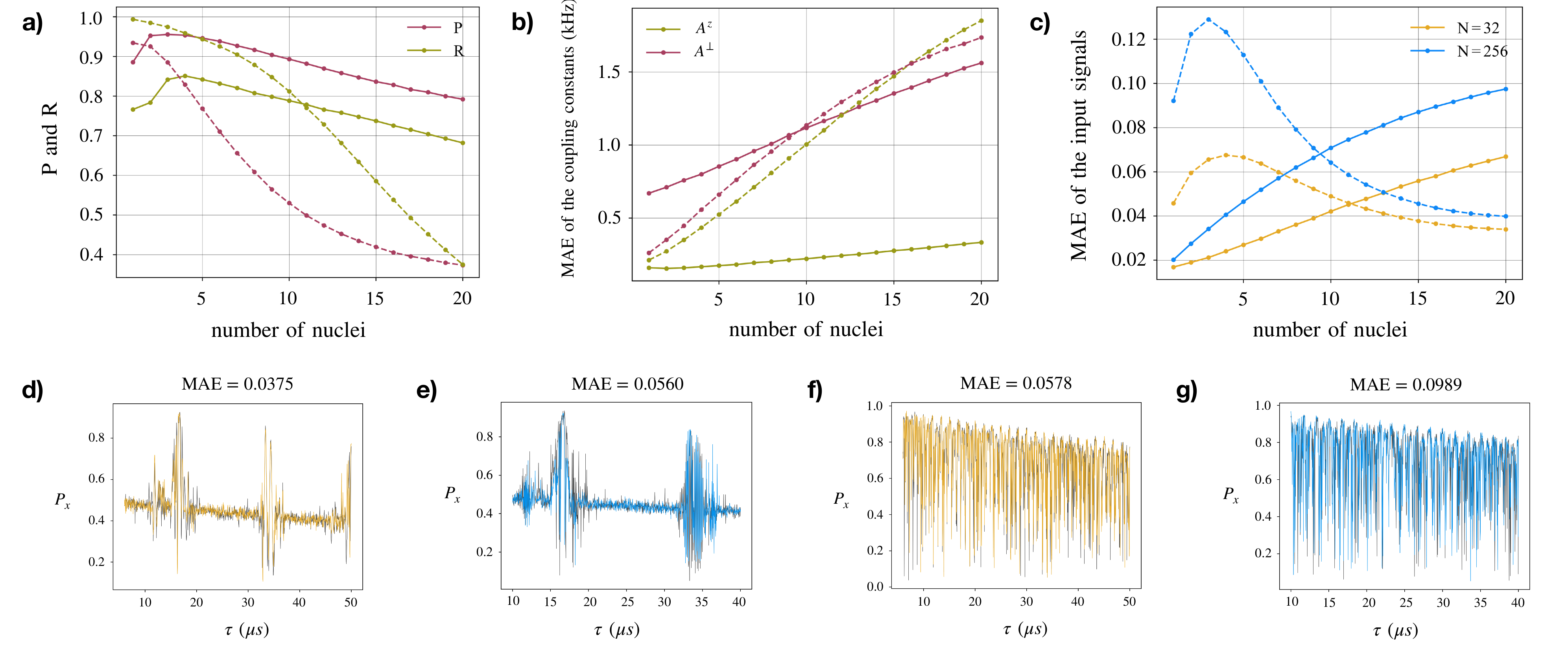}
\caption{(a) Precision ($P$) and recall ($R$) of the model with respect to the number of nuclei present in the true samples. (b) Mean Absolute Error (MAE) between the predicted and the true values of the coupling constants $A^{z}$ (green) and $A^{\bot}$ (red). This metric is calculated exclusively for TPs. (c) In yellow (blue), the MAE between the original and predicted input signals, corresponding to $\rm{N} =32$ ($\rm{N} =256$) pulses. The solid (dashed) lines in (a)-(c) correspond to the high (low) magnetic field scenario. (d)-(g) Signal comparisons from samples containing 16 nuclei. In black, the signals calculated with true nuclei. In yellow (blue), the $\rm{N}=32$ ($\rm{N}=256$) signals obtained with SALI's predicted nuclei. (d) and (e) correspond to the low-field scenario (14 nuclei are detected), and (f) and (g) to high field (all 16 nuclei are detected). More details in~\cite{supplemental}.}
\label{results}
\end{figure*}

In summary, our SALI model operates in two distinct phases: the 1D~$\rightarrow$~2D CNN module and the image post-processing module. Within the neural network module, the 1D CNN block extracts valuable information from the input signals, enabling the representation of hyperfine couplings in a 2-dimensional image through the 2D CNN block. This resulting image serves as a {\it blank canvas} on which the neural network {\it paints} the presence of an initially unspecified number of nuclear spins and estimates their corresponding hyperfine coupling constants. These estimates are subsequently numerically derived through the image post-processing module. Technical details regarding the model are in the SM~\cite{supplemental}, while the SALI code can be requested from the authors.

\emph{Quantifying the model performance and results.--} The values predicted by the model are then compared to the true values using the following procedure. Initially, all nuclei in an image sample are enclosed within bounding boxes: the nuclei in the true sample (``true nuclei'') are bounded in $5\times5$ green boxes and the nuclei in the predicted sample (``predicted nuclei'') are bounded through the image post-processing module in white boxes, as seen in Fig.~\ref{pixel}(d). For each predicted nucleus box, the Intersection over Union (IoU) with each true nucleus box is calculated, defined as the area of overlap over the area of union, ideally being 1; the true nucleus box with the largest IoU is considered the detected true nucleus. 

We employ two standard classification metrics to evaluate the prediction of the number of nuclei: precision, $P = \frac{\rm{TP}}{\rm{TP}+\rm{FP}}$, and recall, $R=\frac{\rm{TP}}{\rm{TP}+\rm{FN}}$. These metrics are computed based on every nucleus within each sample, where true positives (TPs) are correctly detected nuclei, false negatives (FNs) are undetected nuclei, and false positives (FPs) are non-existent nuclei incorrectly detected. High precision (note $P\in[1,0]$) indicates the network's capability to accurately identify existing nuclei without incorporating false ones, while high recall ($R\in[1,0]$) signifies that the network detects a large number of existing nuclei. In Fig.~\ref{results}(a), we present the average of these two metrics calculated for the test subset (15\% of the entire dataset split before training). We have evaluated the performance of coupling constants $(A^z, A^{\bot})$ estimation for the detected nuclei (TPs) using Mean Absolute Error (MAE) between predicted and true values, shown in Fig.~\ref{results}(b), where the MAE remains below $2\ \rm{kHz}$ in all cases.

Finally, we assess the similarity between the original input signals and the ones generated with the complete set of predicted $(A_j^z, A_j^{\bot})$ values by computing the MAE between the original and predicted signals for $\rm{N}=32$ and $\rm{N}=256$ (refer to Fig.~\ref{results}(c)). Notably, although precision and recall decrease as the number of nuclei increases, the predicted signals closely resemble the original signals --as indicated by the low MAEs in Fig.~\ref{results}(c). This conveys the efficacy of SALI in capturing the essential traits of quantum nodes; see Figs.~\ref{results}(d)-(g).

\emph{Conclusions.--}  Employing deep learning models and computer vision algorithms offers a key advantage: rapid and automated detection. Our SALI model achieves large accuracy in predicting hyperfine vectors, effectively handles noisy signals common in experiments, does not require prior knowledge about the number of nuclei in each node, and operates even at low-field conditions. Hence, it serves as a valuable tool for scientists engaged in solid-state platforms, particularly in  quantum sensing and quantum information processing. While our investigation has successfully demonstrated SALI's ability to accurately reproduce input signals, thereby confirming its efficacy, there is potential for further enhancement. One approach to achieve this is by introducing additional signals during the training stage (note, for the sake of simplicity we consider two inputs) obtained with diverse pulse sequences~\cite{casanova15, haase18}. This approach seeks to reinforce the detection process in scenarios characterized by a considerable number of nuclei, particularly in the low-field regime.

\emph{Acknowledgements.--}
E.~G. and J.~C. equally led this work. C.~M.-J. acknowledges the predoctoral MICINN Grant No. PRE2019-088519. J.~C. acknowledges the Ram\'{o}n y Cajal (RYC2018-025197-I) research fellowship, the financial support from Spanish Government via the Nanoscale NMR and complex systems (PID2021-126694NB-C21) project, the Basque Government grant IT1470-22. J.C, E.G., and E.T. acknowledge the ELKARTEK project Dispositivos en Tecnolog\'{i}as Cu\'{a}nticas (KK-2022/00062). Y. B. acknowledges support from the Spanish Government via the project PID2021-126694NA-C22 (MCIU/AEI/FEDER, EU).

\clearpage

\widetext

\begin{center}
\textbf{ \large Supplemental Material: \\ Automatic Detection of Nuclear Spins at Arbitrary Magnetic Fields via Signal-to-Image AI Model}
\end{center}

\setcounter{equation}{0} \setcounter{figure}{0} \setcounter{table}{0}
\setcounter{page}{1} \makeatletter \global\long\def\theequation{S\arabic{equation}}
 \global\long\def\thefigure{S\arabic{figure}}
 \global\long\def\bibnumfmt#1{[S#1]}
 \global\long\def\citenumfont#1{S#1}

\section{Description of the system} \label{math}

We consider a system composed of an NV and $n$ $^{13}\rm{C}$ nearby nuclear spins with Larmor frequency $\omega_{\rm{L}}=\gamma_{\rm{n}}B_z$, such that $\gamma_{\rm{n}} = (2\pi) \times 10.705$ MHz/T, while the magnetic field $B_z$ is aligned with the NV axis ($\hat{z}$). The Hamiltonian that describes this system reads as

\begin{equation}
H_I=\sum_{j=1}^n \omega_j \ \hat{\omega}_j \cdot\vec{I}_j + \frac{1}{2}\sigma_z \sum_{j=1}^n\vec{A}_j \cdot \vec{I}_j + \frac{\Omega}{2}\sigma_{\phi}, 
\label{sH}
\end{equation}

\noindent where $\omega_j \ \hat{\omega}_j = \gamma_{\rm{n}} B_z \  \hat{z} + \frac{1}{2} \vec{A}_j$ are the effective Larmor frecuencies, and $\vec{A}_j=(A_j^{z}, A_j^{\bot})$ are the hyperfine coupling parameters. 

\begin{figure*}[h]
\includegraphics[width=1 \linewidth]{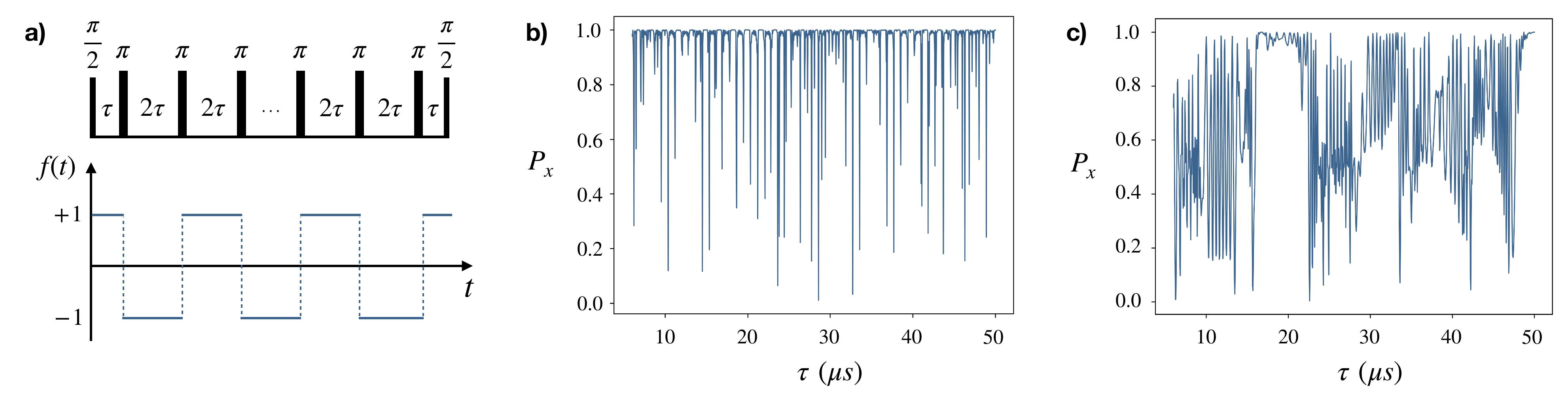}
\caption{(a) Top panel,  CPMG pulse sequence. Initially, a $\pi/2$-pulse is applied to rotate the NV electron spin to the $XY$ plane. Subsequently, $\rm{N}$ $\pi$-pulses are applied, and finally, another $\pi/2$-pulse is applied before the measurement. Bottom panel, $f(t)$ modulation function resulting from the application of the $\pi$-pulses in the CPMG sequence. (b) and (c) $P_x$ calculated at $B_z=0.056 \ \rm{T}$ and $B_z=0.0056 \ \rm{T}$, respectively. Each sequence contains $\rm{N}=32$ $\pi$-pulses, and $P_x$ is sampled $\rm{N}_p=1000$ times in the range $\tau\in[6,\ 50]\ \rm{\mu s}$. In this example, the node contains the NV and 3 nuclei with $(3,75)\ \rm{kHz}$, $(-45,42)\ \rm{kHz}$, and $(23,4)\ \rm{kHz}$.}
\label{scpmgpic}
\end{figure*}

The pulses of the CPMG sequence are applied through a microwave driving represented as the final term in the Hamiltonian from Eq. (\ref{sH}). Specifically, the CPMG sequence involves the application of $\rm{N}$ $\pi$-pulses with an interpulse delay of $2\tau$. In a rotating frame with respect to $H_0 = \frac{\Omega}{2}\sigma_{\phi}$ and applying this pulse sequence, one obtains the following Hamiltonian:

\begin{equation}
H=\sum_{j=1}^n \omega_j \ \hat{\omega}_j \cdot\vec{I}_j + \frac{f(t)}{2}\sigma_z \sum_{j=1}^n \vec{A}_j \cdot \vec{I}_j,
\label{sHi}
\end{equation}

\noindent where the modulation function $f(t)$ is a step function that alternates between $+1$ and $-1$ each time a $\pi$-pulse is applied, as it can be seen in Fig.~\ref{scpmgpic}(a). 

By initializing the NV sensor in an eigenstate of $\sigma_x$ and applying the aforementioned pulse sequence (i.e. CPMG), one can measure the final state of the sensor ($\ket{0}_x$ or $\ket{1}_x$). Repeating this process $\rm{N}_m$ times, one estimates the probability $P_x$ of the state of the sensor being preserved. This probability can be computed for various values of $\tau$ (half the interpulse delay), generating a signal of $P_x$ as a function of $\tau$.

In Fig.~\ref{scpmgpic}(b)-(c), one can observe the $P_x$ performance for distinct values of the magnetic field $B_z$. In particular, in Fig.~\ref{scpmgpic}(b)  ($B_z = 0.056 \rm{T}$) the signal exhibits clear resonance peaks at specific values of $\tau = \frac{k \pi}{2\omega_j}$, for $k$ odd. Fig.~\ref{scpmgpic}(c) shows the behavior of  $P_x$  in a low-field regime ($B_z = 0.0056 \rm{T}$), where these resonances cannot be seen anymore (see mathematical derivation in Supplemental Material from reference \cite{staminiau12}).

\newpage

\section{SALI MODEL}

\subsection{Dataset}

We assessed the performance of the proposed SALI model in two scenarios: one with an NV sensor and up to 20 $^{13}\rm{C}$ nuclei at high magnetic field ($B_z=0.056\ \rm{T}$), and the other with an NV sensor and up to 20 $^{13}\rm{C}$ nuclei at low magnetic field ($B_z=0.0056\ \rm{T}$). The samples were generated based on Eqs. (2-4) in the main text, with 3,600,000 samples for each case. All samples considered a random number of nuclei (up to 20) and random values of $A^{z}$ and $A^{\bot}$ within the ranges $A^{z}\in [-100, \ 100] \ \rm{kHz}$ and $A^{\bot}\in [2, \ 102] \ \rm{kHz}$. Each dataset was divided into training (70\%), validation (15\%), and test (15\%) subsets before network training to prevent bias. Additionally, all three subsets (training, validation, and test) were normalized using the mean value and variance of the training set. The normalization process is expressed as

\begin{equation}
P_x \rightarrow \frac{P_x - \langle P_x \rangle}{\sqrt{\rm{Var}(P_x) + \epsilon}},
\label{normalization}
\end{equation}

\noindent where $\epsilon=0.001$ is added to the denominator to avoid divisions by zero. 

The outputs of the SALI model are the number of nuclei $n$ and the set of coupling parameters $(A_j^{z}, A_j^{\bot})$ between the NV sensor and the nuclear spins. However, it is essential to define an intermediate output for the CNN to supervise the training. Thus, the true output of the network is a 2-dimensional image where nuclei are depicted as Gaussian functions, denoted as $g(x,y) = e^{-\frac{(x-x_0)^2+(y-y_0)^2}{2}}$ (see Fig~\ref{simageproc}(a)). Along the $x$ axis, the perpendicular component of the coupling constant $A^{\bot}$ is represented, while the $y$ axis represents the parallel component $A^{z}$. To portray each nucleus, the exact values of $A^{z}$ and $A^{\bot}$ are taken as the peak positions of the Gaussian function ($x_0$ and $y_0$). The value of the Gaussian at the closest pixel to the peak is used to represent the nucleus, as well as the nearest neighbors (8 pixels) and next-nearest neighbors (16 pixels). As a result, each nucleus is represented by a $5 \times 5$ pixel region in the image. The neural network's output is a $200 \times 100 \ \rm{kHz}$ grid, sized $(200 + 4) \times (100 + 4)$, resulting in pixel sizes of $\Delta A = 1.005 \ \rm{kHz}$ and $\Delta B = 1.01 \ \rm{kHz}$. The ``effective'' output is sized $200 \times 100$, while the 4 extra pixels correspond to a 2-pixel added border to ensure that we can accurately represent nuclei positioned right at the edge of the image, requiring up to 2 additional pixels beyond the image boundary to represent the nearest and next-nearest neighbors. Additionally, we have imposed an upper bound of 1 on pixel values. This constraint addresses situations where two or more nuclei with very similar coupling constants can cause their Gaussian functions to combine and exceed a value of 1 within the same pixel.

This output representation of the neural network allows for the inclusion of an arbitrary number of nuclei and their associated parameters. In terms of the training efficiency of the deep learning model, this representation provides a rich supervisory signal, with each pixel in the density map contributing to the optimization gradients. Additionally, its versatility makes it suitable for various deep learning architectures.

\subsection{Model}

The SALI model comprises two modules: a deep learning network followed by an image processing routine. The model takes two inputs corresponding to signals from CPMG sequences with $\rm{N}=32$ and a $\rm{N}=256$ $\pi$-pulses. The output consists of the set of $(A_j^{z}, A_j^{\bot})$ associated with each of the nuclei predicted by the model. The overall architecture is depicted in Fig. 1 in the main text, featuring a 1D~$\rightarrow$~2D Convolutional Neural Network (CNN) and an image post-processing module.

\subsubsection{1D~$\rightarrow$~2D Convolutional Neural Network (CNN)}

The network architecture comprises 37 layers, organized into three broad blocks:

\begin{itemize}
  \item 1D CNN block: This block consists of two parallel branches designed to independently process each input signal. The input signals undergo two 1-dimensional convolutional layers with 16 and 32 filters, each with a kernel size of 3. Subsequently, these layers are followed by batch normalization, a ReLU layer, and a 1D max-pooling layer that reduces the size of the preceding layer by half. After these layers, a dropout layer with a rate of 0.2 is introduced to mitigate overfitting. The outputs from these two parallel 1D CNN blocks are flattened and then merged into a single array.
  
  \item Fully connected block: The initial dense layer in this block has a size of $(102\cdot52)/2$, serving as the neural network's bottleneck. The subsequent layer is twice the size of the preceding dense layer ($102\cdot52$). Both dense layers are followed by a batch normalization layer prior to the ReLU activation function. A dropout layer with a rate of 0.2 is incorporated between the dense layers. The final dense layer is then reshaped into a $102\times52$ image. This block acts as a transition block between the 1D and 2D CNN blocks.
  
  \item 2D CNN block: This convolutional segment includes a 2-dimensional convolutional layer with 32 filters and a $(3, 3)$ kernel size, followed by a 2-dimensional transposed convolutional layer with 16 filters of the same kernel size. Notably, this last layer employs a stride of 2, effectively doubling the width and height of the input image to the layer. Batch normalization layers are placed between the convolutional layers and the ReLU activation functions, and a dropout layer with a 0.2 rate is added between the convolutional layers. The output of the neural network is a 2D convolutional layer with 1 filter, a $(3, 3)$ kernel size, and a sigmoid activation function that scales each pixel to fall within the range of 0 to 1.
  
\end{itemize}

\subsubsection{Image post-processing}

\begin{figure}[h!]
\includegraphics[width= 1 \linewidth]{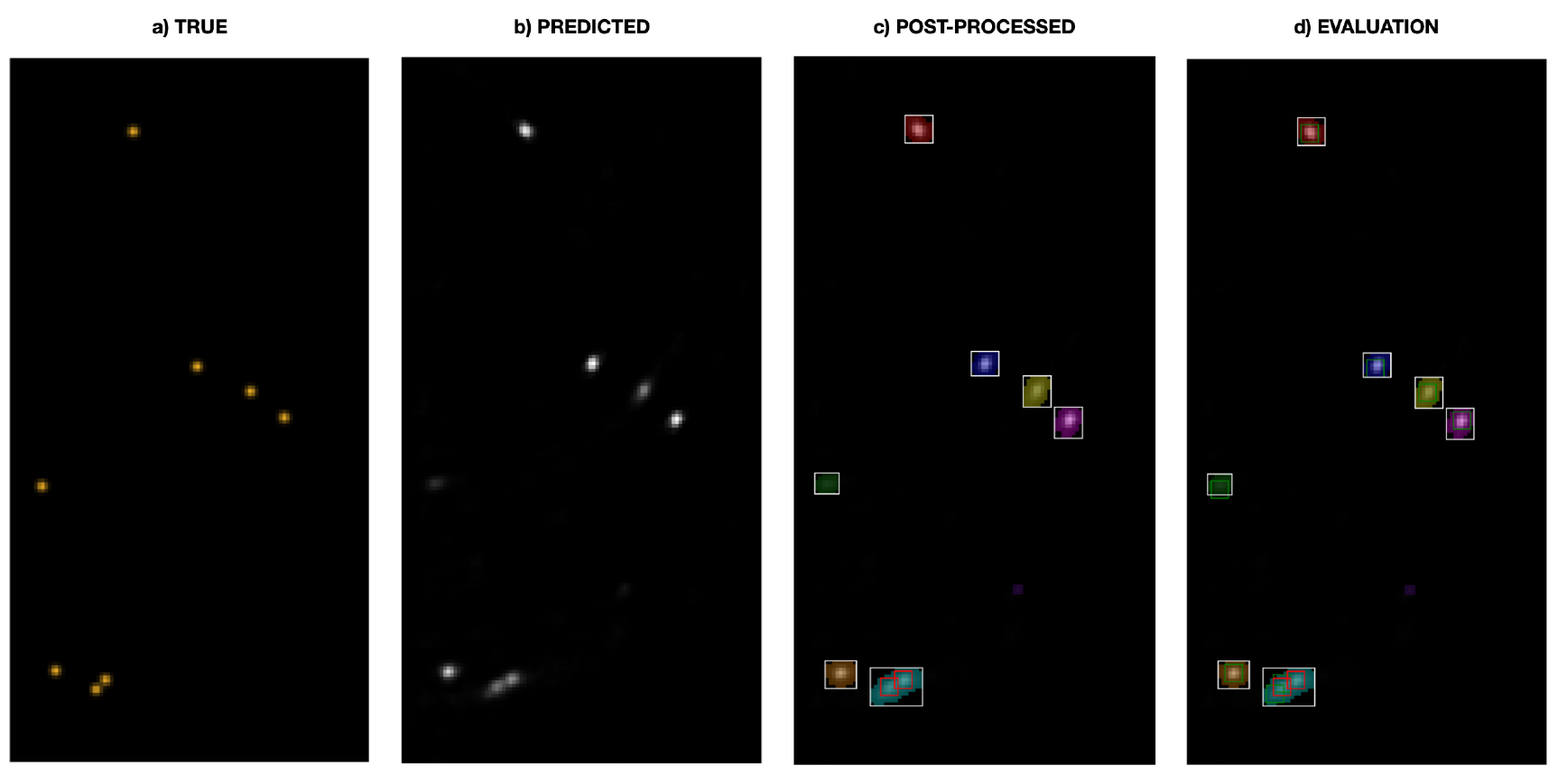}
\caption{(a) True output image taken from the test subset of the low-field case. 8 nuclei are represented in the image, 2 of them closely located. (b) Output image predicted by the neural network. (c) Output image after the post-processing module. The regions corresponding to a single nucleus or, in this particular case, two closely located nuclei, are represented by white boxes. In the region where two nuclei are detected (because there are two maxima), we also see two red boxes, each representing one nucleus. (d) Evaluation of the prediction, depicting the true nuclei in green boxes. All 8 nuclei are correctly predicted.}
\label{simageproc}
\end{figure}

The image post-processing procedure begins with the application of erosion and dilation filters to smooth the image and diminish spurious values in the predicted network outputs (refer to Fig.~\ref{simageproc}(b)). A threshold is then applied to the smoothed image for refinement. Next, adjacent pixels are grouped by connectivity to create clustered objects, followed by the application of an area filter to remove clusters that are deemed too small. These pixel clusters represent regions in the image (depicted as white rectangles in Fig.~\ref{simageproc}(c)). Each region corresponds to a predicted nucleus or a set of ``closely located'' nuclei, meaning their $(A^{z}, A^{\bot})$ values are similar. Each of these regions is examined to identify local maxima. To account for potential local variations in pixel values, a minimum distance of 3 pixels ($\sim3\ \rm{kHz}$) between two maxima in the same region is required to consider them as different nuclei (see Sec.~\ref{spectralselect} for in-depth analysis). If two or more nuclei are detected in the same region, we identify the maxima of each one of them and enclose them in $5\times 5$ red boxes centered around the maxima (see Fig.~\ref{simageproc}(c)). When a single nucleus is detected in a region (i. e. there is just one maximum), the entire region is considered as the detected nucleus, with its centroid serving as the predicted coupling constants of the nucleus.

The neural network outputs can be interpreted as a scaled probability distribution, indicating the potential locations of nuclei. Some predicted nuclei exhibit higher pixel values, signifying the network's confidence in their presence in the sample. Conversely, others may have lower pixel values, indicating lower confidence from the neural network (see Fig.~\ref{simageproc}(b)). This characteristic can be leveraged by adjusting the parameters of the image post-processing (pixel threshold, cluster size...) to achieve a trade-off between precision and recall.

\subsection{Neural network training}

The employed loss function to train the network is Mean Square Error (MSE), calculated across all the pixels in the output grid, and the optimization is performed using the Adam optimizer initialized with a learning rate of 0.001. If there is no improvement in validation loss over 5 epochs, the learning rate decreases by a factor of 0.7. Training is conducted in batches of 64 samples, with dataset shuffling after each epoch. The networks undergo training for a maximum of 250 epochs, employing early stopping with a patience of 20. The model saved from the epoch with the lowest validation loss is considered the final model for testing evaluation. The CNN is implemented using the Keras framework with a TensorFlow backend for training, validation, and testing.

\subsection{Evaluation}

After training, validation, and testing the neural network, image post-processing is applied to the predicted outputs from the test subset to assess the model's performance. Evaluation involves determining the true positives (TP), false positives (FP), and false negatives (FN) for each sample. Specifically, ``true positive" refers to correctly predicted nuclei, ``false positive" indicates the prediction of non-existent nuclei, and ``false negative" signifies that present nuclei were not predicted. To perform this evaluation, all nuclei in an image sample are enclosed within bounding boxes: the nuclei in the true sample (``true nuclei'') are identified in $5\times5$ green boxes, while the nuclei in the predicted sample (``predicted nuclei'') are identified through the image post-processing module in white/red boxes, as described earlier (see Fig.~\ref{simageproc}(d)). For each white/red box (representing a predicted nucleus), we calculate the Intersection over Union (IoU) with each green box (representing the true nuclei), which is defined as the area of overlap over the area of union; ideally, an IoU of 1 indicates a perfect match. The green box with the largest IoU is considered the detected true nucleus. For a nucleus to be considered detected, the IoU between the boxes of the true and predicted nuclei must exceed 0. In addition to tracking true positives (TPs) by measuring IoU, we also compute the Mean Absolute Error (MAE) between the true and predicted coupling constants $A^{z}$ and $A^{\bot}$. Finally, we assess the input signals reconstructed with all predicted nuclei by comparing them to the input signals acquired with the true nuclei employing MAE.

\section{Evaluating spectral selectivity: analyzing model performance in scenarios involving nuclear spins with similar hyperfine couplings} \label{spectralselect}

\begin{figure}[h!]
\includegraphics[width= 1 \linewidth]{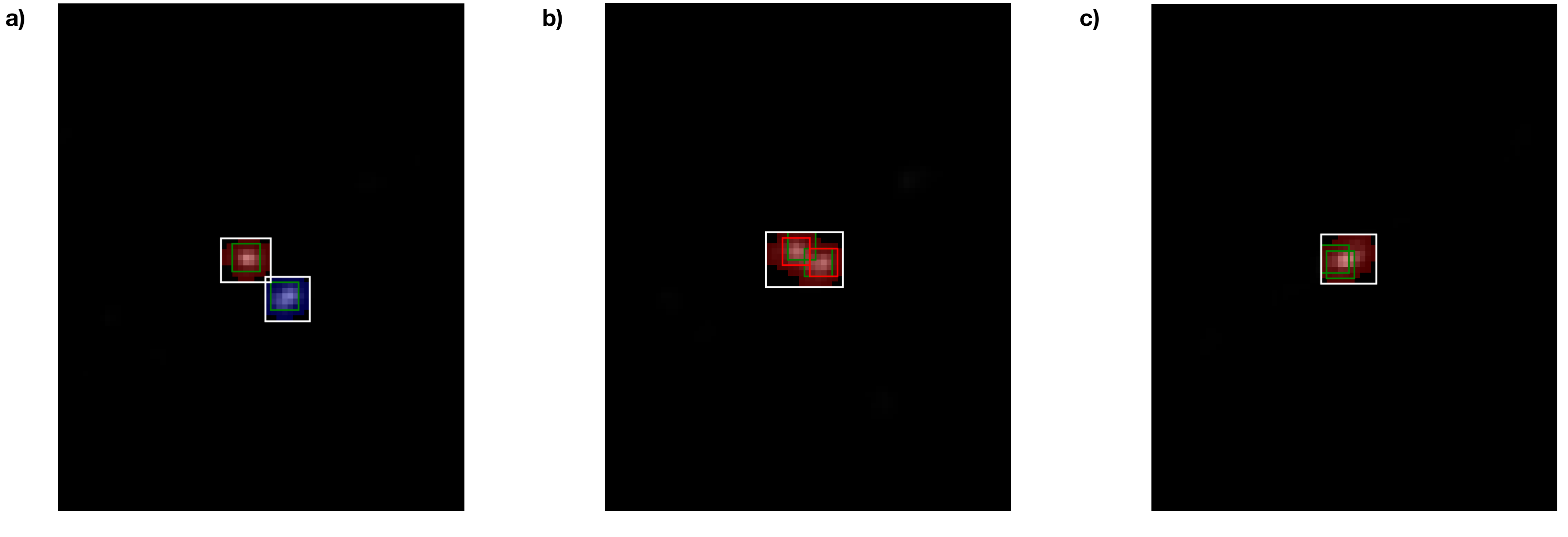}
\caption{Evaluation of the detection of two nuclei with similar coupling constants at low magnetic field. The coupling constants of the first nucleus (depicted in red in (a)) are $(50, 59.77)\ \rm{kHz}$. As the second nucleus approaches the first one, its coupling constants are $(57, 66.77)\ \rm{kHz}$ in (a), $(53, 62.77)\ \rm{kHz}$ in (b), and $(51, 60.77)\ \rm{kHz}$ in (c). In (a) and (b), the two nuclei are detected, occupying different regions in the former and the same region in the latter. Conversely, (c) depicts the detection of a single nucleus.}
\label{2close}
\end{figure}

As mentioned above, the minimum distance required to distinguish two distinct nuclei within the same region is set at 3 pixels during the image post-processing module. Consequently, the theoretical minimum becomes approximately $3\ \rm{kHz}$ for both coupling constants. In Fig.~\ref{2close}, we illustrate the model's performance as the coupling constants of two nuclei with close values progressively approach each other.

\newpage

\section{Testing the robustness of the model against shot-noise} \label{robustnesssection}

\begin{figure}[h!]
\includegraphics[width=0.94 \linewidth]{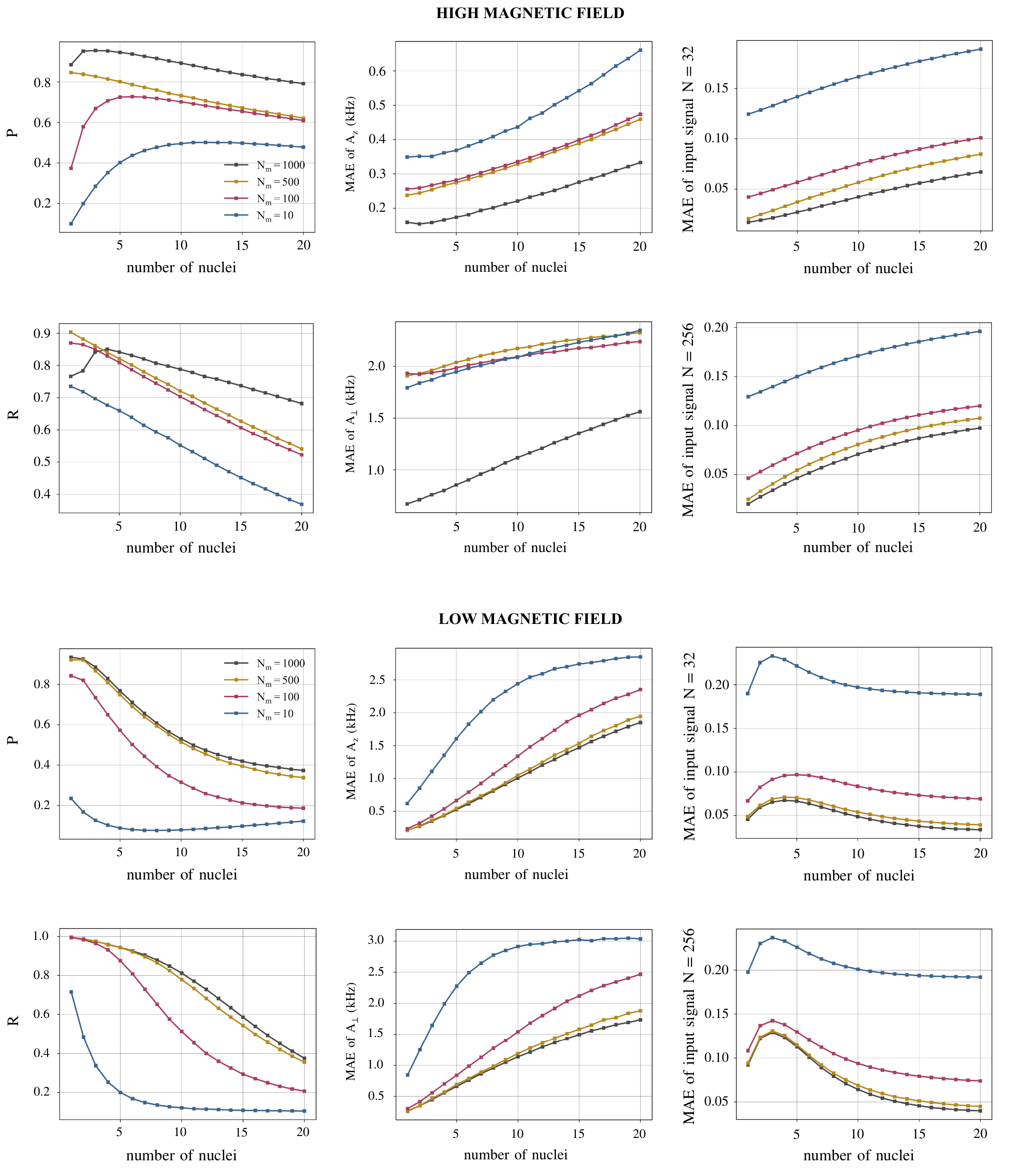}
\caption{Evaluation of the SALI model's performance (initially trained with samples simulated with a shot-noise of $\rm{N}_{\rm{m}}=1000$ measurements) on diverse test sets of 540,000 samples with varying shot-noise levels ($\rm{N}_{\rm{m}}=1000$ --the original test set--, $\rm{N}_{\rm{m}}=500$, and $\rm{N}_{\rm{m}}=100$) to assess its robustness. Plots display multiple metrics, including precision ($P$), recall ($R$), mean absolute error (MAE) for $A^{z}$ and $A^{\bot}$, and the MAE between the original signals (associated with $\rm{N}=32$ and $\rm{N}=256$ $\pi$-pulses in the sequence) and the signals reconstructed with all predicted nuclei. Each line in the plots corresponds to a specific shot-noise level: black for $\rm{N}_{\rm{m}}=1000$ measurements, orange for $\rm{N}_{\rm{m}}=500$, pink for $\rm{N}_{\rm{m}}=100$, and blue for $\rm{N}_{\rm{m}}=10$.}
\label{robustness}
\end{figure}

We conducted simulations with increased noise levels, specifically generating 540,000 samples for $\rm{N}_{\rm{m}}=500$, $\rm{N}_{\rm{m}}=100$, and $\rm{N}_{\rm{m}}=10$ measurements for each datapoint in the input signals, and subsequently tested the model (originally trained with $\rm{N}_{\rm{m}}=1000$) to assess its robustness at both high and low magnetic field. At high magnetic field, although the model experiences a decline in performance with additional noise (specifically for $\rm{N}_{\rm{m}}=500$ and $\rm{N}_{\rm{m}}=100$), the observed trends in metrics remain consistent. At low magnetic field, the model exhibits notable robustness to $\rm{N}_{\rm{m}}=500$ measurements, while displaying a slight performance degradation in the case of $\rm{N}_{\rm{m}}=100$ measurements (see Fig.~\ref{robustness}). This resilience suggests that our model maintains its effectiveness even in the face of increased noise, instilling confidence in its robust capabilities. In the extreme case of $\rm{N}_{\rm{m}}=10$ measurements, the model's performance is noticeably affected by the lack of information at both high and low magnetic field.

\section{Numerical results from the main text}

\begin{table}[h]
\centering
\begin{tabular}{ |C{1.8cm}|C{2.4cm}|C{2.4cm}|C{2.4cm}|C{2.4cm}|C{2.4cm}|C{2.4cm}| }
\hline
\multicolumn{7}{|C{16.2cm}|}{Example: low magnetic field}\\
\hline
Nucleus & True $A^{z}$ (Hz) & Predicted $A^{z}$ (Hz) & True $A^{\bot}$ (Hz) & Predicted $A^{\bot}$ (Hz) & MAE $A^{z}$ (Hz) & MAE $A^{\bot}$ (Hz)\\ 
\hline
1 & -98875.9444 & -98977.6469 & 22447.4441 & 22149.7736 & 101.7025 & 297.6705 \\
2 &   & -91381.1482 &   & 77329.0481 &   &   \\
3 & -82236.4611 & -81407.0352 & 77605.0795 & 79489.1775 & 829.4259 & 1884.0980 \\
4 & -62594.5591 & -62383.3453 & 52589.1205 & 52937.9509 & 211.2138 & 348.8304 \\
5 & -40809.0586 & -40134.6354 & 79724.3165 & 78805.7938 & 674.4232 & 918.5228 \\
6 &   & -31221.3240 &   & 75342.1168 &   &   \\
7 & -6661.2290 &   & 6511.8053 &   &   &   \\
8 & 618.3637 & 215.3625 & 71233.4150 & 71235.2092 & 403.0012 & 1.7942 \\
9 & 22835.2283 & 18848.5981 & 35535.9281 & 37496.4055 & 3986.6302 & 1960.4774 \\
10 & 16596.4429 & 16086.5253 & 41711.0597 & 45149.3541 & 509.9176 & 3438.2945 \\
11 & 15900.9482 & 15723.3536 & 83833.8436 & 81957.4694 & 177.5945 & 1876.3741 \\
12 & 29856.8923 & 31612.6085 & 64685.9245 & 62537.1901 & 1755.7162 & 2148.7344 \\
13 & 34015.0876 & 35494.7985 & 51611.0833 & 47527.5732 & 1479.7109 & 4083.5101 \\
14 & 41004.5308 & 41864.7914 & 35584.3040 & 36961.9774 & 860.2606 & 1377.6733 \\
15 & 77065.1584 &   & 29807.1735 &   &   &   \\
16 &   & 79676.1586 &   & 19732.8844 &   &   \\
17 & 89884.9243 & 92841.0141 & 97399.2709 & 96192.1903 & 2956.0898 & 1207.0806 \\
18 &   & 94358.1544 &   & 49727.2727 &   &   \\
19 & 97146.7497 & 99163.0711 & 89029.5304 & 86725.7760 & 2016.3214 & 2303.7544 \\
20 & 97293.2625 & 97487.4372 & 37515.6950 & 36848.4848 & 194.1747 & 667.2102 \\
\hline
\end{tabular}
\caption{Numerical values of the true and predicted coupling constants $(A^z, A^{\bot})$, as well as the corresponding MAEs of the example from Fig.~3(d)-(e) in the main text. 14 nuclei are detected (TPs) out of the 16 true nuclei. 4 nuclei are over-predicted (FPs).}
\label{tableexamplelowfield}
\end{table}

\begin{table}[h]
\centering
\begin{tabular}{ |C{1.8cm}|C{2.4cm}|C{2.4cm}|C{2.4cm}|C{2.4cm}|C{2.4cm}|C{2.4cm}| }
\hline
\multicolumn{7}{|C{16.2cm}|}{Example: high magnetic field}\\
\hline
Nucleus & True $A^{z}$ (Hz) & Predicted $A^{z}$ (Hz) & True $A^{\bot}$ (Hz) & Predicted $A^{\bot}$ (Hz) & MAE $A^{z}$ (Hz) & MAE $A^{\bot}$ (Hz)\\ 
\hline
1 & -89149.1218 & -89391.4015 & 58528.9955 & 59126.8238 & 242.2796 & 597.8283 \\
2 & -89250.2135 & -88996.2634 & 17861.7656 & 17565.9156 & 253.9501 & 295.8500 \\
3 & -72472.3346 & -72437.9473 & 76943.9533 & 76196.5106 & 34.3873 & 747.4427 \\
4 & -71864.3395 & -71922.1106 & 43449.0304 & 45797.3485 & 57.7711 & 2348.3181 \\
5 & -67856.0332 & -67967.1083 & 62388.9558 & 63707.9890 & 111.0751 & 1319.0332 \\
6 & -43283.2279 & -43177.4256 & 36742.4227 & 37291.3753 & 105.8023 & 548.9526 \\
7 & -34559.2966 & -34673.3668 & 72799.0305 & 72707.0707 & 114.0702 & 91.9598 \\
8 & -25741.1784 & -25642.2826 & 41337.3828 & 41100.0937 & 98.8958 & 237.2891 \\
9 & -21229.8613 & -20136.9824 & 31394.1066 & 32611.2804 & 1092.8789 & 1217.1738 \\
10 & -5100.0355 & -4781.0481 & 63366.8872 & 65145.7431 & 318.9874 & 1778.8560 \\
11 &   & -3661.1630 &   & 49835.4978 &   &   \\
12 & 38751.9170 & 38627.5640 & 30000.2914 & 28444.7756 & 124.3530 & 1555.5158 \\
13 & 41270.5583 & 41909.5477 & 70786.8143 & 68707.0707 & 638.9895 & 2079.7436 \\
14 & 46661.3024 & 47180.3462 & 47310.4908 & 46332.2110 & 519.0437 & 978.2798 \\
15 & 69741.6539 & 69770.4207 & 29130.9627 & 27272.3312 & 28.7668 & 1858.6315 \\
16 & 76707.7658 & 76469.3030 & 40118.2734 & 40647.3430 & 238.4628 & 529.0696 \\
17 & 87940.3129 & 88358.4590 & 40976.3481 & 40236.5320 & 418.1461 & 739.8161 \\
\hline
\end{tabular}
\caption{Numerical values of the true and predicted coupling constants $(A^z, A^{\bot})$, as well as the corresponding MAEs of the example from Fig.~3(f)-(g) in the main text. All 16 nuclei are detected (TPs). 1 nucleus is over-predicted (FPs).}
\label{tableexamplehighfield}
\end{table}

\begin{table}[h]
\centering
\begin{tabular}{ |C{1.5cm}|C{1.8cm}|C{1.8cm}|C{2.2cm}|C{2.2cm}|C{2.6cm}|C{2.6cm}| }
\hline
\multicolumn{7}{|C{14.7cm}|}{Model metrics: low magnetic field}\\
\hline
\# of true nuclei & $ \rm{P} $ & $ \rm{R} $ & MAE $A^{z}$ (Hz) & MAE $A^{\bot}$ (Hz) & MAE input signal N=32 & MAE input signal N=256\\ 
\hline
1 & 0.9341 & 0.9939 & 211.8066 & 261.2769 & 0.0458 & 0.0921 \\
2 & 0.9255 & 0.9848 & 272.2131 & 350.5561 & 0.0595 & 0.1224 \\
3 & 0.8853 & 0.9744 & 349.7452 & 446.5646 & 0.0656 & 0.1290 \\
4 & 0.8289 & 0.9589 & 434.0356 & 557.3587 & 0.0676 & 0.1233 \\
5 & 0.7684 & 0.9435 & 524.2099 & 660.1902 & 0.0666 & 0.1129 \\
6 & 0.7104 & 0.9255 & 613.1820 & 761.0398 & 0.0638 & 0.1010 \\
7 & 0.6562 & 0.9047 & 710.9949 & 863.9525 & 0.0598 & 0.0891 \\
8 & 0.6087 & 0.8789 & 808.9583 & 955.3167 & 0.0560 & 0.0792 \\
9 & 0.5648 & 0.8479 & 909.0699 & 1048.2445 & 0.0523 & 0.0708 \\
10 & 0.5301 & 0.8121 & 1003.8220 & 1135.0094 & 0.0489 & 0.0642 \\
11 & 0.4987 & 0.7702 & 1100.3608 & 1212.2242 & 0.0459 & 0.0586 \\
12 & 0.4738 & 0.7285 & 1202.8350 & 1294.2948 & 0.0433 & 0.0543 \\
13 & 0.4525 & 0.6815 & 1290.1381 & 1365.7002 & 0.0412 & 0.0508 \\
14 & 0.4348 & 0.6342 & 1384.2604 & 1432.3049 & 0.0394 & 0.0479 \\
15 & 0.4194 & 0.5855 & 1470.0856 & 1496.6571 & 0.0378 & 0.0456 \\
16 & 0.4055 & 0.5383 & 1562.0888 & 1558.5923 & 0.0365 & 0.0437 \\
17 & 0.3960 & 0.4925 & 1641.0533 & 1605.6893 & 0.0355 & 0.0423 \\
18 & 0.3879 & 0.4518 & 1718.2080 & 1657.2404 & 0.0348 & 0.0412 \\
19 & 0.3797 & 0.4120 & 1788.7925 & 1692.8111 & 0.0343 & 0.0404 \\
20 & 0.3734 & 0.3749 & 1851.7677 & 1735.7796 & 0.0340 & 0.0399 \\
\hline
\end{tabular}
\caption{Numerical values from Fig.~3(a)-(c) in the main text: low magnetic field. These results are calculated separately depending on the number of nuclei present in the true samples.}
\label{tableresultslowfield}
\end{table}

\begin{table}[h]
\centering
\begin{tabular}{ |C{1.5cm}|C{1.8cm}|C{1.8cm}|C{2.2cm}|C{2.2cm}|C{2.6cm}|C{2.6cm}| }
\hline
\multicolumn{7}{|C{14.7cm}|}{Model metrics: high magnetic field}\\
\hline
\# of true nuclei & $ \rm{P} $ & $ \rm{R} $ & MAE $A^{z}$ (Hz) & MAE $A^{\bot}$ (Hz) & MAE input signal N=32 & MAE input signal N=256\\ 
\hline
1 & 0.8855 & 0.7663 & 159.3502 & 669.6763 & 0.0169 & 0.0202 \\
2 & 0.9524 & 0.7837 & 154.4154 & 710.4753 & 0.0190 & 0.0275 \\
3 & 0.9556 & 0.8417 & 158.6041 & 758.5075 & 0.0212 & 0.0341 \\
4 & 0.9536 & 0.8509 & 166.2072 & 799.9738 & 0.0241 & 0.0406 \\
5 & 0.9462 & 0.8419 & 173.8719 & 853.4260 & 0.0270 & 0.0465 \\
6 & 0.9383 & 0.8317 & 181.4522 & 902.5522 & 0.0297 & 0.0519 \\
7 & 0.9271 & 0.8207 & 193.6377 & 957.8667 & 0.0331 & 0.0571 \\
8 & 0.9166 & 0.8076 & 201.5162 & 1007.4918 & 0.0361 & 0.0620 \\
9 & 0.9041 & 0.7983 & 212.6115 & 1067.0834 & 0.0390 & 0.0663 \\
10 & 0.8933 & 0.7885 & 221.0919 & 1116.6486 & 0.0421 & 0.0709 \\
11 & 0.8817 & 0.7786 & 232.2747 & 1164.1576 & 0.0452 & 0.0746 \\
12 & 0.8696 & 0.7659 & 242.2465 & 1208.1121 & 0.0478 & 0.0779 \\
13 & 0.8582 & 0.7581 & 252.0024 & 1261.3507 & 0.0505 & 0.0811 \\
14 & 0.8472 & 0.7477 & 263.8876 & 1305.2804 & 0.0534 & 0.0844 \\
15 & 0.8364 & 0.7373 & 276.0918 & 1352.9696 & 0.0559 & 0.0871 \\
16 & 0.8283 & 0.7257 & 286.0759 & 1394.1018 & 0.0580 & 0.0895 \\
17 & 0.8168 & 0.7153 & 296.8244 & 1439.9926 & 0.0606 & 0.0917 \\
18 & 0.8098 & 0.7040 & 309.8346 & 1482.3191 & 0.0628 & 0.0938 \\
19 & 0.8000 & 0.6931 & 321.4237 & 1525.2497 & 0.0649 & 0.0958 \\
20 & 0.7921 & 0.6820 & 333.2698 & 1561.6735 & 0.0669 & 0.0975 \\
\hline
\end{tabular}
\caption{Numerical values from Fig.~3(a)-(c) in the main text: high magnetic field.}
\label{tableresultshighfield}
\end{table}

\end{document}